\documentclass[aps,preprint]{revtex4}
\usepackage{graphicx,amssymb}
\begin{document}
\setcounter{page}{1}
\title{ $\Delta$H = $\Delta$B region in a Volume Defect-Dominating \\Superconductor}
\author{H. B. \surname{Lee}}
\author{G. C. \surname{Kim}}
\author{H. J. \surname{Park}}
\author{D. \surname{Ahmad}}
\author{Y. C. \surname{Kim}}
\email{yckim@pusan.ac.kr}
\thanks{Fax: +82-51-513-7664}
\affiliation{Department of Physics, Pusan National University, Busan 46241, Korea}

\begin{abstract}
It has been generally accepted that the diamagnetic property of type II superconductor decreases after H$_{c1}$. On the other hand, 
 we  found that (Fe, Ti) particle-doped MgB$_2$ specimens have  a $\Delta$H = $\Delta$B region in the M-H curves, which is the region that the increase of a magnetic induction
is as much as the increase of an applied magnetic field. Here we study whether this phenomenon was only confined to 
 (Fe, Ti) particle-doped MgB$_2$  superconductor, whether there is a theoretical basis, and why it does not  appear in other superconductors.
 The cause of the $\Delta$H = $\Delta$B region was  the pinning phenomenon of defects in the superconductor and it only occurs  in volume defect-dominating superconductors. Widths of the $\Delta$H = $\Delta$B region along the number of defects and H$_{c2}$
  were calculated, and compared with the experimental results. We hypothesized that pinned fluxes have to be depinned from the defect and move into an inside of a superconductor regardless $\Delta$G$_{defect}$ if the distance between fluxes pinned at the volume defect is equal to that of H$_{c2}$.  The region means that the fluxes that have penetrated into the inside of a superconductor are  pinned preferentially on the volume defects over the entire specimen before the general behavior.
 
\end{abstract}



\maketitle

Although it is clear that superconductors have a flux pinning effect, the exact mechanism is not completely understood \cite{Schuster, Ghorbani, Geim, MingXu}. 
 All superconductors have flux pinning effects  because they have defects even if defects are few. Most of type II superconductors have shown that the diamagnetic property of the superconductor decreases gradually after H$_{c1}'$ (not  H$_{c1}$), which is  defined as the field showing the maximum diamagnetic property.
On the other hand, we observed a phenomenon in (Fe, Ti) particle-doped MgB$_2$ specimens that do not show the general behavior after H$_{c1}'$, which is the existence of  a $\Delta$H = $\Delta$B region in the magnetization-applied field (M-H) curves. This is an unusual phenomenon that has not reported in other superconductors. 
 
The flux pinning phenomenon is  caused mainly by defects in the superconductor.  
Generally, defects in the superconductor contain volume defects (such as general volume defects,  precipitators, inclusions
and columnar defects, etc.), planar ones (such as grain boundaries, twin boundaries and stacking
faults plane, etc.), and line ones (such as dislocations). Although they all belong to a family of defects, a role
difference 
for flux pinning is considerable. In the case of volume
defects,  
 pinned fluxes are difficult to escape from defects 
 except when a force balance
(F$_{pinning}$ = F$_{pickout}$) is broken; hence, they are called strong pinning sites. 

On the other hand,
weak pinning sites, such as planar defects and line defects, are entirely different
from the strong pinning sites in the 
flux pinning mechanism.
The grain boundaries (GBs), which have
relatively lower pinning energy caused by planar characteristic, are connected to each other in the entire
specimen as a planar defect. Therefore,   
fluxes
pinned on the GB move easily along the GBs.  In addition, because the total area as a defect is large, they have significant importance in the overall 
flux pinning effects. A superconductor dominated by planar defects in the flux pinning effect can be called a planar defect-dominating superconductor.  High T$_c$ superconductor (HTSC) bulks are associated in this category \cite{Shimizu, Tkaczyk, Senoussi, Suryanarayanan}. 

The depinning, which means the phenomenon that magnetic fluxes escape from the defect, will be considered by two ways.  The one is pick-out depinning, which is the depinning that fluxes pinned on the volume defect are depinned together. The other is leak-out depinning, which is the depinning that fluxes pinned on the volume defect are depinned one by one. The former is the depinning by the force balance of fluxes pinned at the defect and the latter is depinning through grain boundaries connected on volume defect because grain boundaries  does not only pin the fluxes but also leak out fluxes pinned at volume defects. 
Regarding dislocations, the penetration of fluxes through them is not too difficult as planar defect-dominating superconductors do because dislocations as a line defect are also interconnected throughout the specimen,.   
 Worked NbTi superconducting wires are associated in this category \cite{ bean, Ghosh}.  Therefore, planar and line-defect dominating superconductors appear superficially to follow the general behavior.   

  MgB$_2$, which was made by a synthetic method at high temperatures also has grain boundaries, but most of them are low angle ones due to their fabricating characteristic \cite{Kang, Dou, Eom, Jin, Canfield}. 
  Hence it has significantly fewer weak links than HTSC bulks produced by a solid state reaction method. Therefore, it can be called a volume defect-dominating superconductor. 
  MgB$_2$ has been known as a superconductor which field dependence is weak, 
 but a definite effect can be obtained by  doping artificial defects because it is a volume defect-dominating superconductor \cite{Dou, Dou1,Lee, Lee1, 
Lee2}.

Pure MgB$_{2}$ and  (Fe, Ti) particle-doped MgB$_{2}$ specimens for this study were synthesized using a non-special atmosphere synthesis (NAS) method \cite{Lee}. 
 All specimens which had been synthesized at 920$^o$C  for 1 hour were cooled in air, but 
  5 wt.\% (Fe, Ti) particle-doped MgB$_{2}$ specimen, which had shown prominent results, 
 underwent two different cooling processes. One was cooled in air and the other was quenched in water.  Figure \ref{fig2} (a) presents the NAS method for MgB$_2$ and Fig. \ref{fig2} (b) shows a photograph of the method. 
Figure \ref{fig2} (c) is a photograph of (Fe, Ti) particles, which are slightly far from sphere and Fig. \ref{fig2} (d) shows (Fe, Ti) particles  present in MgB$_2$. 
 The radius of the particles is rather irregular, and the average radius of them is 163 nm. 

 \section{Results }
  \subsection{ A diamagnetic property increase and the confirmation of the $\Delta$H = $\Delta$B region in experiments}
 Fluxes would penetrate into the superconductor in flux quantum form over H$_{c1}$ \cite{Little}. The diamagnetic property of the superconductor decreases gradually after the maximum property
  and this phenomenon continues to H$_{c2}$. This is true if there are no defects, which are pinning sites in the superconductor. However,  real superconductors which have defects behave differently.  Fluxes, having penetrated into the superconductor, are pinned at the defects near the surface and the diamagnetic property increases rather than that of H$_{c1}$. We call it H$_{c1}'$, which represents the field of the maximum diamagnetic property in a real superconductor. 
 
In planar defect-dominating superconductors and line defect-dominating superconductors, the diamagnetic property of H$_{c1}'$ did not make a large difference from that of H$_{c1}$ if there are no volume defects. It is caused by the fact that the small volume of an individual defect induces a weak pinning force. 
Hence,  the increase of the diamagnetic property at H$_{c1}'$ is small. In particular, they are interconnected; thus, they allow well for  flux penetration. Therefore, it appears to follow the general behavior superficially and there is no the $\Delta$H = $\Delta$B region in the M-H curves. 
 
 On the other hand, volume defect-dominating superconductors show distinctly different behavior. The pinning effect is strong due to their relatively larger volume, and the most important thing is that they are not interconnected with each other. Therefore, they continue to pin fluxes until their pinning limits. They would act as another barrier to prevent the fluxes from penetrating into the superconductor over H$_{c1}$. Thus,  the diamagnetic property of the volume defect-dominating superconductors certainly increases. As shown in all the M-H curves except for pure MgB$_2$ in Fig. \ref{fig3}, 
 a linear region ends  about 600 Oe, which means perfect diamagnetism. After that, they show a slight decrease in slope. This behavior means that the fluxes penetrated into the superconductor are pinned at defects near the surface and cannot  move easily into the specimen.
  Therefore, the diamagnetic property of the specimen continues to increase even though H$_{c1}$ has passed.

 Figure \ref{fig3} (a) presents M-H curves of pure MgB$_2$ and 5 wt.$\%$ (Fe, Ti) particle-doped MgB$_2$ that were air-cooled and measured at 5 K.  The M-H curve of pure MgB$_2$ used as reference. It is clear that  the $\Delta$H = $\Delta$B region is observed after  H$_{c1}'$ 
  in 5 wt.$\%$ (Fe, Ti) particle-doped MgB$_2$. The width of the region  can be disputed, but it is definite that the M-H curve of the specimen forms a $\Delta$H = $\Delta$B region from the H$_{c1}'$ to 8 kOe. After flux jump, it continues to show the $\Delta$H = $\Delta$B region up to 15 kOe. And it shows  gradual decrease of diamagnetic properties, which are the $\Delta$H $>$ $\Delta$B region over 15 kOe. 
 
Figure \ref{fig3} (b) presents the M-H curve of 5 wt.$\%$ (Fe, Ti) particle-doped MgB$_2$ that was water-quenched and measured at 5 K. Generally, the water-quenching method is used to refine  the grains by impeding the growth rate of grains or to induce a rapid phase transformation (e.g.: martensite transformation). 
  In current experiments, it was used to increase the angle between the  grains of MgB$_2$ and refine the grains due to the rapid cooling rate. This treatment has the purpose of providing further opportunities for the fluxes that have been pinned at volume defects to leak out through the grain boundary. Thus, this procedure can reduce the stress of the concentration of fluxes on the volume defects.
 
Therefore,  the $\Delta$H = $\Delta$B region in the figure is formed up to 20 kOe in a wide view, even though there was a small flux jump. As shown in these two figures, it is reasonable that the $\Delta$H = $\Delta$B region of the 5 wt.$\%$ (Fe, Ti) particle-doped MgB$_2$ specimen is from  H$_{c1}'$ to a point between 15 kOe and 20 kOe.
  Figure \ref{fig3} (c) presents the M-H curve  measured at 10 K on the air-cooled 5 wt.$\%$ (Fe, Ti) particle-doped MgB$_2$ specimen and (d), (e) and (f) in Fig. \ref{fig3} are ones measured at 5 K with different doping concentrations. It is clear that  the $\Delta$H = $\Delta$B region is observed in all specimens except for the 1  wt.\% doped specimen.
  
 \subsection{Pinned fluxes movement and the basis of  $\Delta$H = $\Delta$B region}
A previous study reported  that the flux quanta pinned at a defect move with a bundle and hop from one pinning site to another  \cite{Gurp, Bonevich}. 
 If the distance between the volume defects is wide enough, the fluxes that are pinned at the defect move when the force balance is broken (F$_{pinning}$ $<$ F$_{pickout}$), which is based on $\Delta$G$_{defect}$ and a repulsive force between the flux quanta. On the other hand, when the distance between the volume defects is short,   fluxes pinned at the defect would move into an inside of the superconductor by a different mechanism. This means that when  a volume defect of the superconductor reach the limit value of pinned fluxes, they have to be depinned from the defect and move into an inside of the superconductor regardless of  $\Delta$G$_{defect}$

If a volume defect existing near the surface of the superconductor pins fluxes 
and 
they are blocked from moving into an inside of the superconductor until defect's pinning limit,
  the free energy density of a spherical  defect in the superconductor can be expressed as 
{\setlength\arraycolsep{2pt}
{\setlength\arraycolsep{2pt}
\begin{eqnarray}
\Delta G_{super}-\Delta G_{nor}=\frac{H^2}{8\pi} \Rightarrow
\Delta G_{defect} = -\frac{H^2}{8\pi}\times\frac{4}{3}\pi r^3  
\end{eqnarray}
 where $H$  is the applied field and r is the radius of the defect.
According to the equation, $\Delta$G$_{defect}$ is dependent on  the external field $H$ when r is constant. The flux quanta pinned at the defect can move into an inside of the superconductor when they are at F$_{pinning}$ $<$ F$_{pickout}$ state, and they will be pinned again at another defect in front of them. 
 It is necessary to increase H in order for the fluxes to be depinned from the defect and move into an inside of the superconductor. 
 If $H$ is increased, $\Delta$G$_{defect}$  in the superconductor becomes larger. Therefore, a stronger $H$ will be needed for penetrating fluxes into an inside of the superconductor. 
 
Although a diamagnetic property of the superconductor increases due to the pinning phenomenon, it does not increase continuously. 
 There must be a limit of the pinned fluxes that brakes this premise as shown in the experiment.  
 We considered the basis of this limit to be the minimum distance between the pinned fluxes at the defect. This is because  the neighborhood around  the defect which have pinned fluxes is no longer a superconducting state when the minimum distance between  the fluxes pinned at the defect is less than that of H$_{c2}$; thus, there is no pinning effect anymore.
 
The reason for creating  a $\Delta$H = $\Delta$B region in the M-H curve is originated from the flux pinning limit of a volume defect in the superconductor.  
 The $\Delta$H = $\Delta$B region is formed in the M-H curve because the defects are filled with flux quanta step by step from the surface to the center of the superconductor when the defects have a pinning limit of flux quanta according to their radius. 
  Figure \ref{fig5} (a) and (b) show the flux pinning limit of defects for having the same and different radius, respectively. Both are superconductors with four defects along y-axis and nine defects along x-axis. The difference between the two is the particle size of defects,  which is uniform in (a) and the average particle size in (b) is the same as that of (a). When the flux quantum lies in y-axis and moves along x-axis direction, it is natural that the $\Delta$H = $\Delta$B region is formed in (a) because the fluxes coming from the outside of the defect are pinned at the defect and move if a defect exceeds its flux  pinning limit. 
 
 On the other hand, Figure \ref{fig5} (b) shows slightly different behavior. Fluxes pinned at a small pinning site moves first because the flux pinning limit is low, and fluxes pinned at a larger defect move later. When many fluxes that have been pinned at larger defect are depinned from the defect, they move together; thus, there is a high possibility of flux jump. The 5 wt.$\%$ (Fe, Ti) particle-doped MgB$_2$ is an example of   unevenness of defects,  as shown in Fig. \ref{fig3} (a). It has approximately  8000$^3$ defects in 1 cm$^3$ of MgB$_2$, of which the radius is 163 nm on average. 
  In this situation, a single quantum flux in a superconductor would be simultaneously pinned at 8000 defects on average. 
 Although there are some fluxes pinned at a defect that move first and some fluxes at another defect that move later, the ability to pin fluxes on average is similar to the counterpart when it is observed as a whole specimen.
 Therefore, there is no problem in forming the $\Delta$H = $\Delta$B region. 
 
\subsection{ Calculations for a flux pinning limit  of a defect and the width of  $\Delta$H = $\Delta$B region} 

Assuming that volume defects are spherical, their size is constant, and they are arranged regularly in a superconductor, a superconductor of  1 cm$^3$ has m$^3$ volume defects. The maximum number of flux quanta that can be accommodated at a spherical defect of radius r 
 in a static state is
\begin{eqnarray}
n^2 = \frac{\pi r^2}{\pi (\frac{d}{2})^2}\times P = (\frac{2r}{d})^2\times P
 \end{eqnarray} 
 where r, d and P is the radius of defects, a distance between quantum fluxes and filling rate which is $\pi$/4 when they have square structure, respectively (see Fig. \ref{fig5} (c) and (d)).
If the radius of a defect  is 163 nm, the maximum number of quantum fluxes that can be pinned by the defect is approximately 45$^2$ at 0 K in the static state because the distance (d) is 6.43 nm when H$_{c2}$ is 50 T (H$_{c2}$ = $\Phi_0/d^2$) \cite{Tinkham}.
We thought that quantum fluxes had a square structure rather than a triangular one when they were pinned at the defect \cite{Abrikosov}. 

Therefore, the magnetic induction
B can be expressed as
 \begin{eqnarray}
B = n^2m_{cps}m\Phi_0
 \end{eqnarray} 
where $n^2$, m$_{cps}$, m, and $\Phi_0$ are the number of quantum  
fluxes pinned at a defect,  the number of defects which are in the vertically closed packed state, 
the number of
defects with pinned fluxes from the surface  
to the center of the superconductor, and flux quantum, respectively.  m$_{cps}$ is explained in Fig. \ref{fig5} (e) and (f).
 m$_{cps}$ is the minimum number of defects when the penetrated fluxes into the superconductor are completely pinned. This conversion was introduced to calculate the number of flux quanta which are pinned on defects of a plane because the fluxes between defects can penetrate into the superconductor without pinning if  defects are  arranged regularly like a lattice as shown in Fig. \ref{fig5} (e). 
The conversion is much closer to reality  because  defects are arranged randomly in a real superconductor. If 8000$^3$ defects are in 1 cm$^3$ superconductor, as described in the experiment, there are approximately 8000$^2$ defects in a plane. Therefore, there are almost no penetrating fluxes without pinning. Thus,  the total number of flux quanta pinned on the defects of a plane perpendicular to the flux-moving direction are $n^2m_{cps}$.
  
 Hence, the magnetization M  is
 \begin{eqnarray}
B = H +4\pi M \Rightarrow  M = \frac{B-H}{4\pi} = \frac{ n^2m_{cps}m\Phi_0 - H}{4\pi}
 \end{eqnarray} 
Therefore, a width of the $\Delta$H = $\Delta$B region is
 \begin{eqnarray}
\Delta H = H - H_{c1}' = n^2m_{cps}m\Phi_0 - 4\pi M -  H_{c1}'
 \end{eqnarray} 
where $H_{c1}'$ is the field showing the first maximum diamagnetic property in the superconductor.
If the radius of defects is fixed, $n^2$ and $m_{cps}$ are also fixed. Therefore, the width of the $\Delta$H = $\Delta$B region is  dependent only on the m.
A calculated width of the $\Delta$H = $\Delta$B region  along a number of defect  is shown in Fig. \ref{fig6} (a) when the radius of a defect is 163 nm and H$_{c2}$ is 50 T. As shown in the figure, the width of the  $\Delta$H = $\Delta$B region increases with increasing number of pinning sites except for over-doping. 

One of  the important factors calculating the width of the $\Delta$H = $\Delta$B region is what is H$_{c2}$ of a superconductor.  H$_{c2}$ is a fundamental property according to the material of a superconductor, but it is inferred from the indirect method at a low temperature because it has difficulty in being  measured directly. For example, H$_{c2}$ of MgB$_2$ varies from a theoretical value of 64 T to experimental one of approximately 20 T   \cite{Buzea, sung, Sologubenko, Poole}. 
  The calculated width of the $\Delta$H = $\Delta$B region along H$_{c2}$ variation is shown in Fig. \ref{fig6} (b)  when the radius of a defect is 163 nm and there are 8000$^3$ defects in a 1 cm$^3$ superconductor, which is equivalent to 5 wt.\% (Fe, Ti) particle-doped MgB$_2$. 
  
When the width of the $\Delta$H = $\Delta$B region was calculated with H$_{c2}$ = 50 T, it reasonably matches the experimental results as   shown in Fig. \ref{fig6} (a) and (b), which were converted from 5 K to 0 K (the width of the $\Delta$H = $\Delta$B region was conservatively determined to be 1.3 T at 5 K in the 5 wt.\% specimen, thus it will be 5.2 T if expanded by 1 cm because the thickness of measured specimen is 2.5 mm). The experimental result is rather higher than the theoritical one in the figures, low purity of boron (96.6 \%) caused volume defects of which radius is 1$ \mu$m on average (SMFig. 7).  It is determined that the cowork of (Fe, Ti) particles with them make the result. 
Figure \ref{fig6} (c) shows the flux penetration method based on the general behavior \cite{bean2} and (d), (e) and (f) show the flux penetration method based on the existence of the $\Delta$H = $\Delta$B region. They indicated that the fluxes penetrated into the superconductor are pinned preferentially on the volume defects over the entire specimen before the general behavior. The width of the $\Delta$H = $\Delta$B region increases with increasing number of volume defects, and the width of the region is narrow if the number of volume defects are few or too many.

 \section{Discussion}
The presence of  $\Delta$H = $\Delta$B region  is of great importance in practical applications of the superconductor. Consider, for example, the case of using superconductors in magnetic levitation train. Superconductors showing the general behavior are difficult to use  diamagnetic property up to the maximum. When the train levitates and moves, there will be up and down vibrations, which will bring in more magnetic fields on the superconductor. When the magnetic field is applied beyond the field which produces the maximum diamagnetic property, there is a fear that the train may fall to the bottom because diamagnetic property will be reduced. On the other hand, superconductors with $\Delta$H = $\Delta$B region have no problem even when using the maximum diamagnetic property because the maximum diamagnetic property is maintained in a considerable region if there is no flux jump.

To solve a weak magnetic field dependence of MgB$_2$, many researchers have doped a variety of materials and achieved considerable results \cite{Shi, Silva, Chengduo}. However, despite the improved field dependence in high field, there were still a lot of flux jumps in low field. 
 Therefore, it was easy that  the $\Delta$H = $\Delta$B region in MgB$_2$ specimens was overlooked. In our experiments, we didn't recognize the region owing to flux jump in 5\% doped specimen. However, we suspected the diamagnetization point after flux jump (the point of 1.5 T in Fig. \ref{fig3} (a)), which was too much higher, and confirmed the region after quenching the specimen in water, which lowered the flux jump (Fig. \ref{fig3} (b)).

The essence of this communication is as follows. Fluxes that have been penetrated into the superconductor 
 are  pinned preferentially on volume defects over the entire specimen if it is  a volume defect-dominating superconductor.  This is because  fluxes pinned on the defects are bent like a bow; thus, unpinned ones are difficult to exist without pinning on defects due to the repulsive force between fluxes and the irregular distribution of defects.  Since the volume defect reach its pinning limits when the external field exceeded H$_{c1}'$, the internal fluxes (B) increase as much as  the external field (H) increases. Therefore, when the superconductor is is dominated by volume defects, $\Delta$H = $\Delta$B region is first formed after H$_{c1}'$ and the $\Delta$H  $>$ $\Delta$B region is formed later in the M-H curve. 
 
On the other hand, it might be hard to accept that flux pinning on defects cause a larger diamagnetic property than that of H$_{c1}$. However, this is a common phenomenon because there is no material having no defects. It is rather natural to explain that planar and line defect-dominating superconductors follow the general behavior is due to the interconnectivity of the defects. A typical example of increasing the diamagnetic property by flux pinning is the fishtail effect. The fishtail effect is often observed in superconducting single crystal (SC), particularly in HTSC SCs. There are many opinions about the cause of the fishtail effect, but there is some consensus that it is due to the pinning phenomenon  \cite{Radzyner, Daeumling}.

 One of the important features of volume defect-dominating superconductor (VDS) is the flux jump. 
  If pinned fluxes on volume defects do not leak out through grain boundaries, the volume defects will pin fluxes to their pinning limit.  
 In addition, they move together when they are picked out from the defect; thus, flux jump can occur if they are many. Moreover,  
  because  diamagnetic property of the VDS is always higher than that of the pure state of superconductor and the superconductor with volume defects are pinned from the surface, the fluxes pinned on defects are always under pressure that they may penetrate into an inside of the superconductor. 
 This is the reason that the flux jump  occurs well in MgB$_2$  synthesized at high temperature. 
 
  The main reason that a $\Delta$H = $\Delta$B region has not been reported so far is considered to be low density of volume defects and the lack of a proper VDS like MgB$_2$. In addition, the thickness of the measured specimen also influence considerable effect on the width of $\Delta$H = $\Delta$B region. 
   Though the density of volume defect is meaningful,  the region cannot be observed owing to its thickness if a measured specimen is thin. The number of pinning sites decreases as the thickness of the specimen becomes thinner, 
  thus the region do not appear to be distinguished level. 
  
 As shown in Fig. \ref{fig3}, the $\Delta$H = $\Delta$B region does not appear when the density of the volume defect is low (Fig. \ref{fig3} (d)) and the region is too short to recognize it when the density of the volume defect is high (Fig. \ref{fig3} (f)).   
 The important thing in the $\Delta$H = $\Delta$B region is the number of volume defects as well as the density of them. Under the same density of defects, it is difficult to observe the region if the volume defects are large and a few, whereas observation of the region is possible if they are small and many. In our experiments, we have compared the width of the regions according to the density of volume defect using of which radius is  163 nm on average, showing that 5 wt.\% doped specimen had the widest region and width of the region tends to decrease if the volume defects are denser or sparser.
  
Other examples of VDS are unworked NbTi and melt-texture growth (MTG) specimens of HTSC. In the case of NbTi, it is difficult to find an experiment of a correlation with the volume defect because the focus for the increase of the pinning effect was more on the line defects. However, unworked NbTi can be classified as a VDS owing to its flux jump \cite{Chabanenko}.
On the other hand,  melt-texture growth (MTG)  was introduced to eliminate the weak links of HTSCs.  
 One of the distinct features of MTG  is that flux jump occurs frequently like MgB$_2$ \cite{Xing}. The fact that the flux jump, which was not observed in the HTSC specimen prepared by the solid-phase reaction method, frequently occurs in a HTSC specimen prepared by the MTG method means that the dominating flux pinning mechanism has changed from planar defects to volume defects.  Because MTG  has a higher concentration of impurities, 
an extensive literature search on the M-H curves of MTG was performed and two papers were found \cite{Hsu, Muller}. The considerable width of the $\Delta$H = $\Delta$B region has been formed after H$_{c1}'$ in these papers. Therefore, 
we had clear confirmation that the phenomenon that $\Delta$H = $\Delta$B region appears is not to be confined to MgB$_2$, especially this experiment, but to be common in volume defect-dominating superconductors.
 
 The $\Delta$H = $\Delta$B region was demonstrated by experiments of (Fe, Ti) particle-doped MgB$_2$ specimens that are a volume defect-dominating superconductor. And we represented a theoretical base of the phenomenon and compared with experiment results. Moreover, we found a superconductor in a literature that show a $\Delta$H = $\Delta$B region in MTG HTSC, and confirmed the generality of the phenomenon. 
It is considered that the behavior of the superconductor is based on the flux pinning limit of the volume defects regardless of $\Delta$G$_{defect}$. In addition, it was also emphasized that superconductors should be classified not as materials but as defects in order to understand  the flux pinning phenomena properly. 
 There is no defect-free material, and it is proper to interpret the phenomenon of superconductivity based on this point. The $\Delta$H = $\Delta$B region has the basis of the flux pinning phenomenon, which appears ahead of the general behavior in a volume defect-dominating superconductor. 

\section{Method}
 The starting materials were Mg (99.9\% powder), B (96.6\% amorphous powder) and (Fe, Ti) particles. The mixed Mg and B stoichiometry, and  (Fe, Ti) particles were added by weight. They were finely ground and pressed into 10 mm diameter pellets. The (Fe, Ti) particles were ball-milled for several days, and the average radius of the (Fe, Ti) particles was approximately 0.163 $\mu$m. 
   On the other hand, an 8 m-long stainless- steel (304) tube was cut into 10 cm pieces. One side of the 10 cm-long tube was forged and welded. The pellets and excess Mg were placed in the stainless-steel tube. The pellets were annealed at 300$^o$C  for 1 hour to make them hard before inserting them into the stainless-steel tube. The other side of the stainless-steel tube was also forged. High-purity Ar gas was put into the stainless-steel tube, and which was then welded. All of the specimens were synthesized at 920$^o$C  for 1 hour. The field and temperature dependence of magnetization were measured using a MPMS-7 (Quantum Design). During the measurement, sweeping rates of all specimens were equal for the same flux-penetrating conditions.

$\\$

 $\bf{Acknowledgements}$\\  
The authors would like to thank Dr. B. J. Kim of PNU for careful discussion and Dr. L. K. Ko and Dr. D. Y. Jeong of KERI for experimental supports.  \\ 

$\bf{Author}$ $\bf{Contributions}$\\ This communication was conceived by H. B. Lee, experimented by H. B. Lee and G. C Kim,  written by H. B. Lee, G. C. Kim, D. Ahmad and Y. C. Kim. H. J. Park did mathematical calculation. \\ 

$\bf{Competing }$ $\bf{ Interests}$\\ The authors declare no competing interests.


\newpage
\begin{figure}
\begin{center}
  \includegraphics*[width=
  16cm]{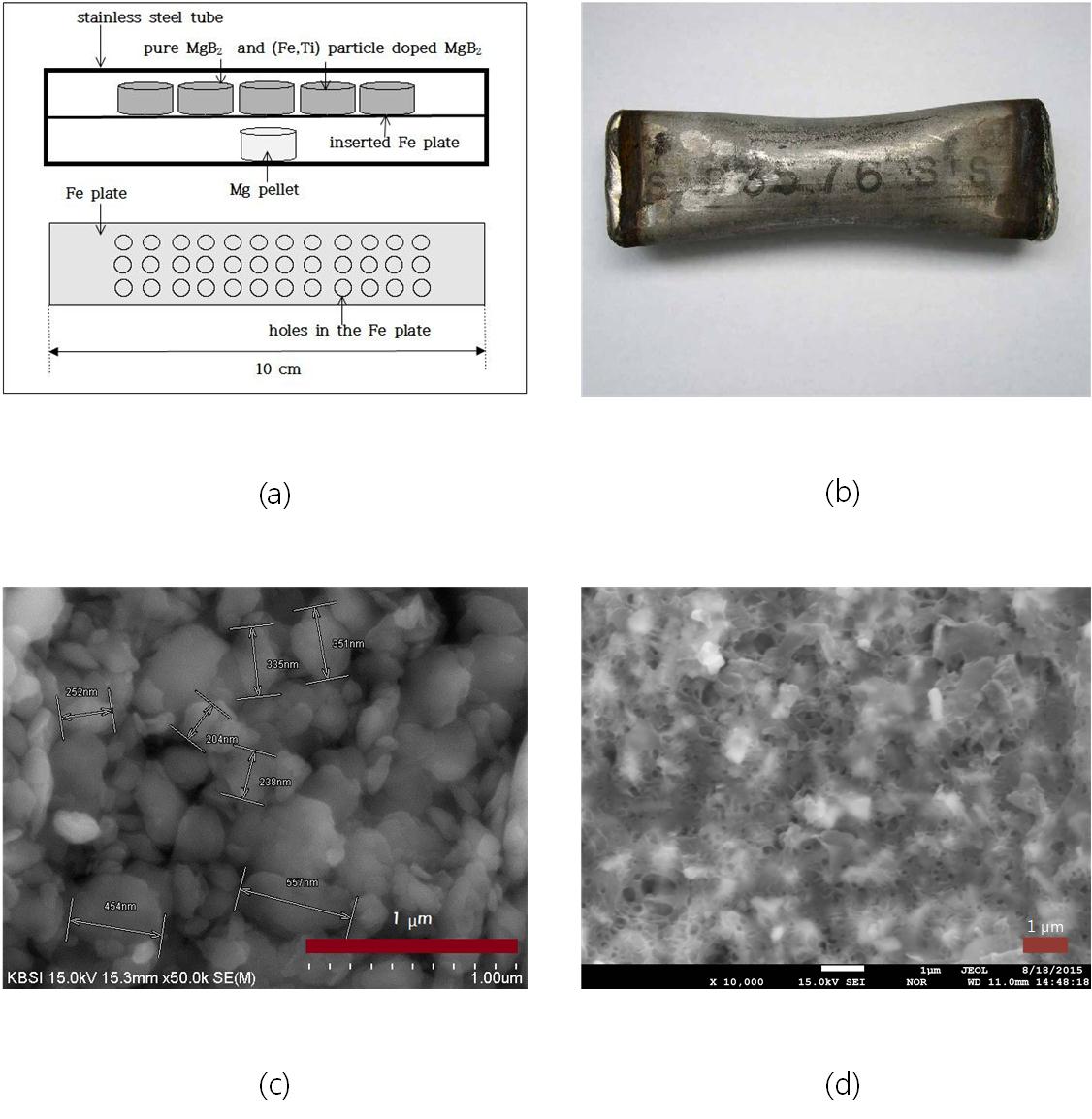}
   \end{center}
\caption{ Non-special atmosphere synthesis (NAS) method for MgB$_2$  and (Fe, Ti) particles for the experiment.  (a): Schematic representation of the non-special atmosphere synthesis (NAS) method. (b): Photograph of the specimen for the NAS method. (c): Photograph of (Fe, Ti) particles before doped in MgB$_2$, which were ball-milled for several days.  (d): A photograph of 25 wt.$\%$ (Fe, Ti) particle-doped MgB$_2$, which was taken by field emission scanning electron microscope (FE-SEM).  The white bright ones in the MgB$_2$ base are doped (Fe, Ti) particles.}
 \label{fig2}
 \end{figure}

\newpage
\begin{figure}
\begin{center}
  \includegraphics*[width=
  14cm]{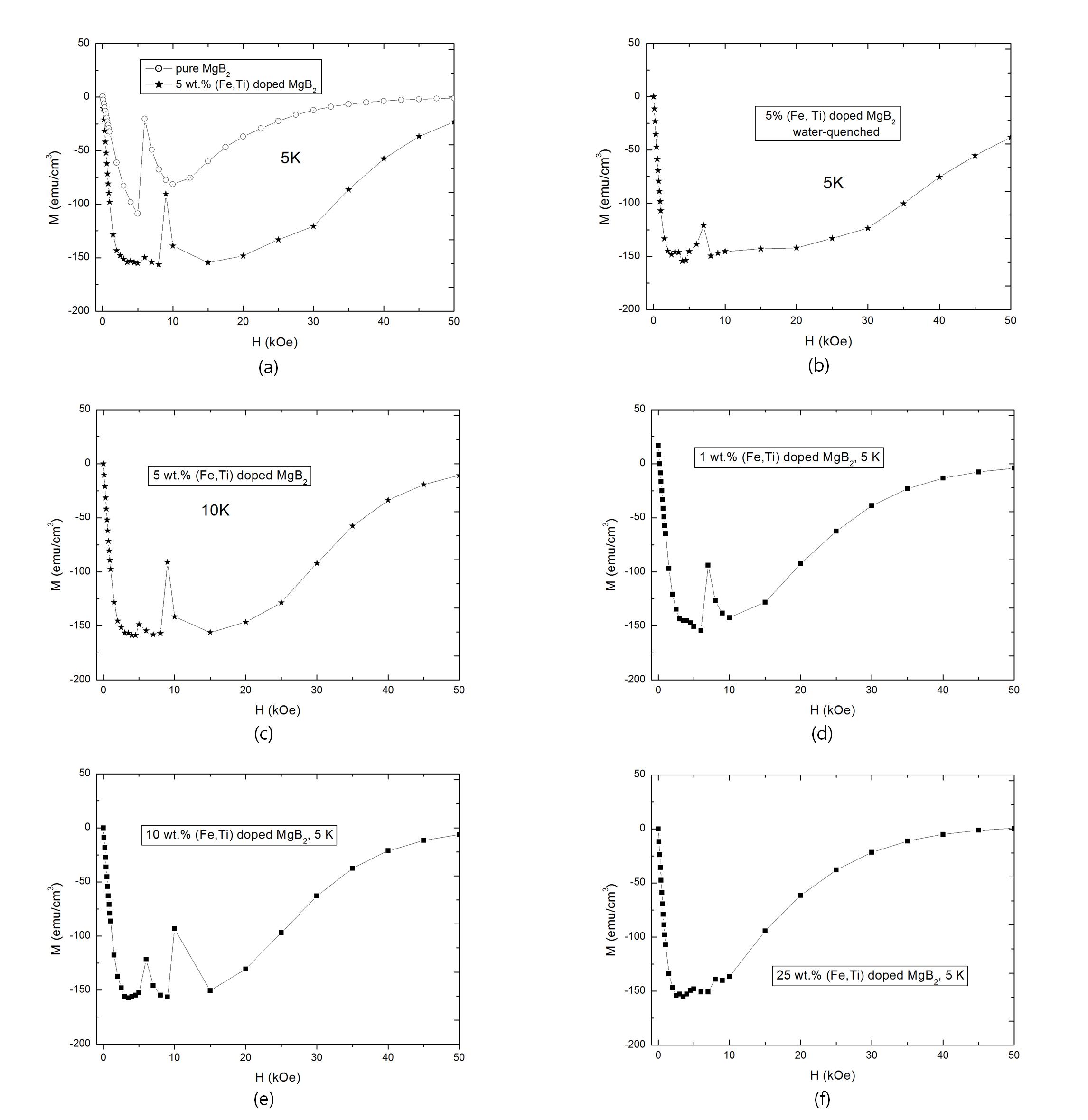}
   \end{center}
\caption{  Field dependence of magnetization for pure MgB$_2$  and (Fe, Ti) particle-doped MgB$_2$ (M-H curves). (a): Field dependence of magnetization for pure MgB$_2$  and 5 wt.$\%$ (Fe, Ti) particle-doped MgB$_2$. Specimens were air-cooled and measured at 5 K. (b): Field dependence of magnetization for 5 wt.$\%$ (Fe, Ti) particle-doped MgB$_2$, which was water-quenched and  measured at 5 K. (c): Field dependence of magnetization for 5 wt.$\%$ (Fe, Ti) particle-doped MgB$_2$, which was air-cooled but measured at 10 K. (d): Field dependence of magnetization for 1 wt.$\%$ (Fe, Ti) particle-doped MgB$_2$, which was air-cooled and measured at 5 K. (e): Field dependence of magnetization for 10 wt.$\%$ (Fe, Ti) particle-doped MgB$_2$, which was air-cooled and measured at 5 K. (f): Field dependence of magnetization for 25 wt.$\%$ (Fe, Ti) particle-doped MgB$_2$, which was air-cooled and measured at 5 K.  Full versions are shown in Supplementary Materials}
 \label{fig3}
 \end{figure}

\newpage
 \begin{figure}
\begin{center}
\includegraphics*[width=11cm]{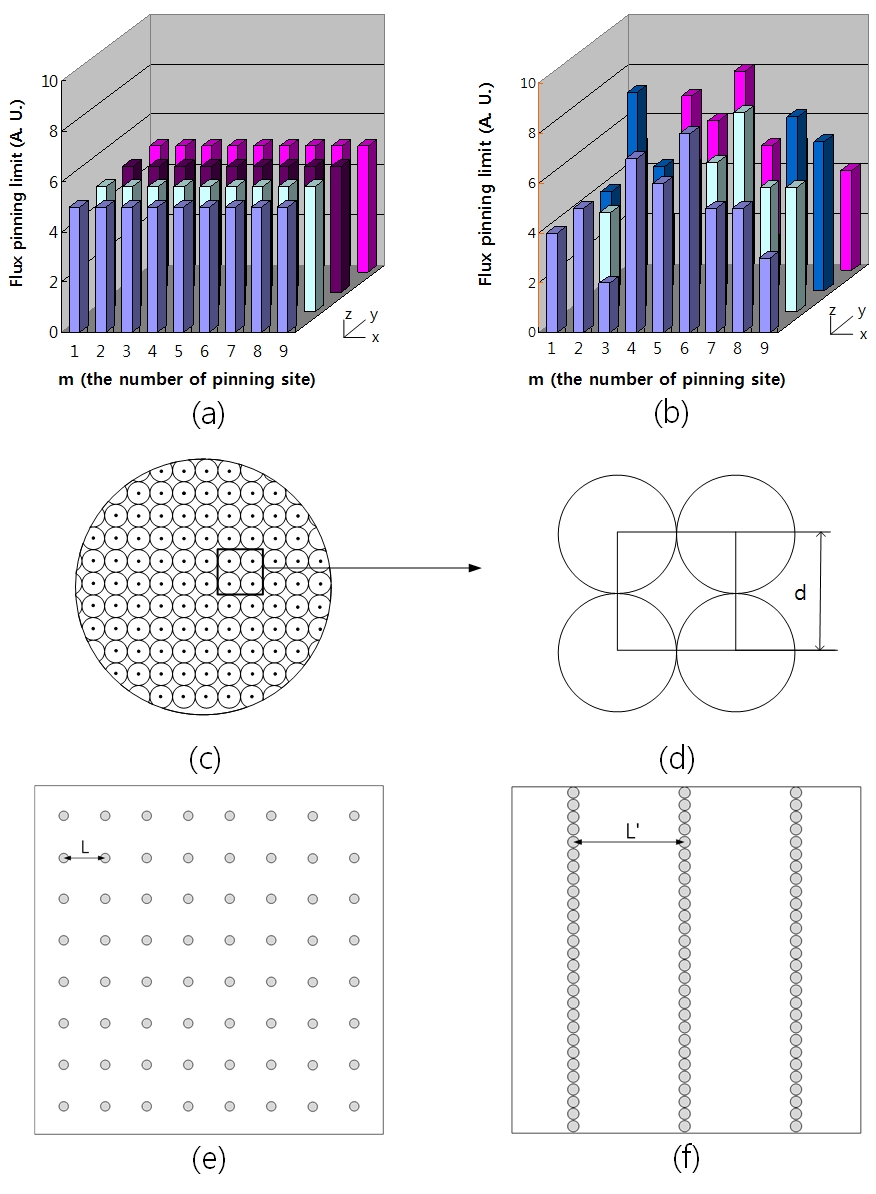}
\end{center}
\caption{Flux pinning limit of defects, filling rate calculation and the definition of  m$_{cps}$. (a): Flux pinning limit of defects when they have same radius and regular arrangement. (b): The flux pinning limit of defects when they have different radii and a regular arrangement.  (c): Shape of the maximum fluxes pinned at the defect, which is cut off the center of the defect and assumed to be spherical. 
(d): The definition of d 
(e): Ideal arrangement of defects. There is a possibility that  fluxes  are not pinned at defects if the fluxes lie on the y axis and move along the x axis. (f): The definition of m$_{cps}$.  The m$_{cps}$ is the number of defects which are a vertically closed packed state of defects.  
The defect arrangement in (e) needs to change to that in the (f) for calculating B in the superconductor for not having any flux quantum penetrating into the superconductor without pinning.
}
\label{fig5}
\end{figure}

\newpage
 \begin{figure}
\begin{center}
\includegraphics*[width=13cm]{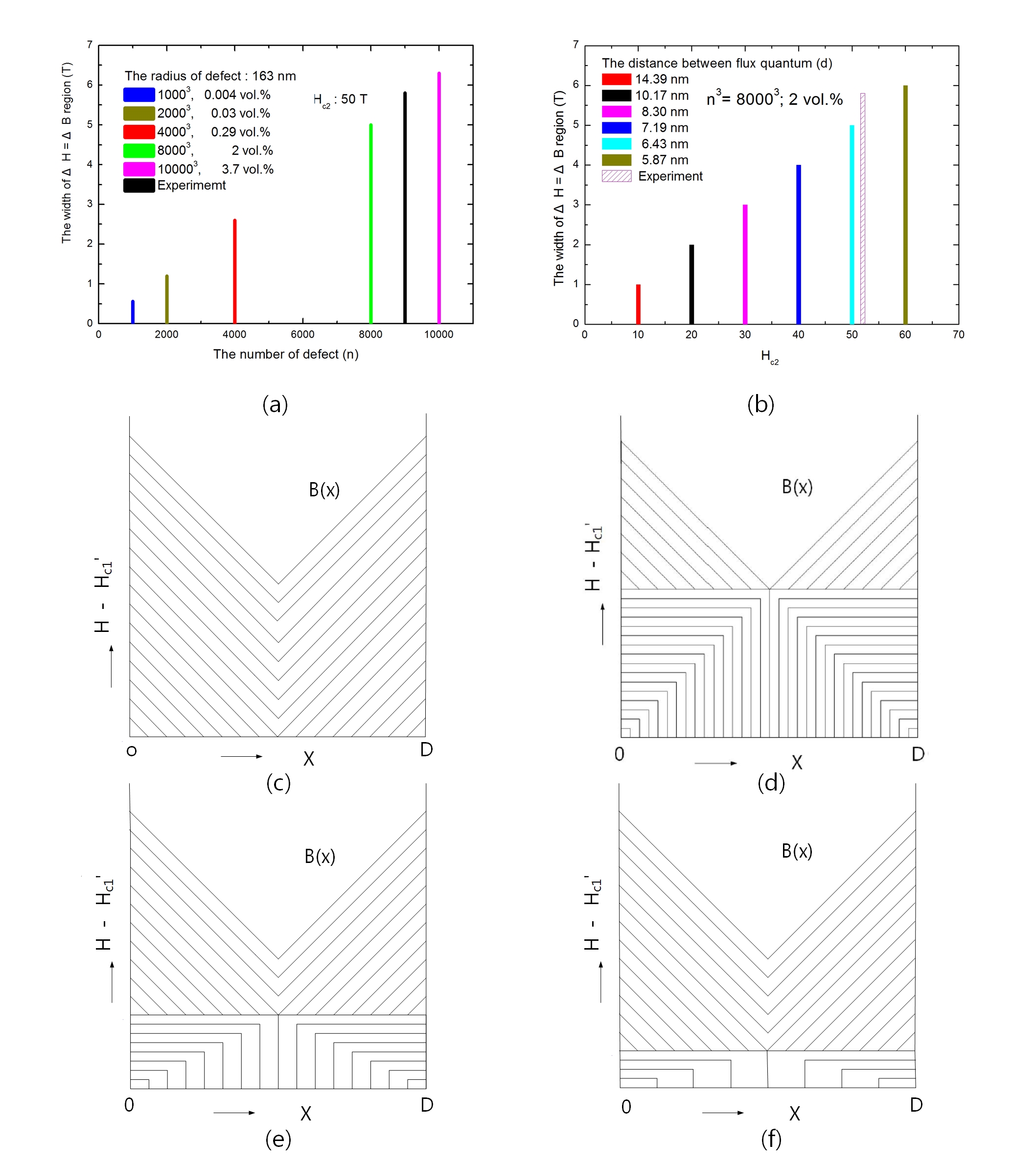}
\end{center}
\caption{Calculated width of the $\Delta$H = $\Delta$B region and flux penetration method compared to Bean Model. (a): Calculated width of the $\Delta$H = $\Delta$B region along the number of defects in a superconductor. (b): Calculated width of the $\Delta$H = $\Delta$B region along the upper critical field (H$_{c2}$) of a superconductor. (c): Flux penetration method which are based on the general behavior 
\cite{bean2}.   (d): Flux penetration method when the  superconductor has a good pinning condition in a volume defect-dominating superconductor. Fluxes penetrated into the superconductor are  pinned  on volume defects from existing ones around the surface of the superconductor. 
(e): Flux penetration method when  the superconductor has a proper pinning condition in a volume defect-dominating superconductor.  (f): Flux penetration method when  the superconductor has a poor pinning condition in a volume defect-dominating superconductor.}
\label{fig6}
\end{figure}



\end{document}